\documentclass[]{spie}  
\usepackage[]{graphicx}
\usepackage{url}

\title{BoA: a versatile software for bolometer data reduction} 

\author{Fr\'ed\'eric Schuller\supit{a,b}
\skiplinehalf
\supit{a}European Southern Observatory, Chile \\
\supit{b}Max-Planck-Institut f\"ur Radioastronomie, Germany
}

\newcommand{\mic}{~$\mu$m}

\begin{document} 
  \maketitle 

\begin{abstract}
     Together with the development of the Large APEX Bolometer Camera (LABOCA)
     for the Atacama Pathfinder Experiment (APEX), a new data reduction package
     has been written. This software naturally interfaces with the telescope
     control system, and provides all functionalities for the reduction,
     analysis and visualization of bolometer data. It is used at APEX for
     real time processing of observations performed with LABOCA and other
     bolometer arrays, providing feedback to the observer. Written in an
     easy-to-script language, BoA is also used offline to reduce APEX
     continuum data. In this paper, the general structure of this software
     is presented, and its online and offline capabilities are described.
\end{abstract}


\keywords{Data reduction; bolometers; sub-millimeter}

\section{INTRODUCTION}
\label{sec:intro}  

The Atacama Pathfinder Experiment (APEX) is a novel 12~m single dish
sub-millimeter (submm) telescope, operating from the Chanjnantor Plateau
in Chile
since 2006\cite{guesten2006}. It is equipped with heterodyne receivers
covering frequencies of 230 GHz to 1.4 THz (wavelengths 200 microns
to 1.3~mm), and several large arrays of bolometers: the APEX-SZ
camera\cite{ref-aszca} at a wavelength of 2~mm,
LABOCA (870\mic) and SABOCA (350\mic).
More bolometer arrays are expected to come in the near future.

The Large APEX Bolometer Camera\cite{siringo2009} (LABOCA) consists
of 295 bolometer detectors operating at 870\mic. They are arranged
in a hexagonal pattern, with two-beam spacing between individual
detectors, resulting in a field of view of 11.4$'$ in diameter.
As part of the development of LABOCA by the Bolometer Group in the
Max-Planck-Institut f\"ur Radioastronomie (MPIfR), a new data
reduction software has been written for processing bolometer
data acquired at APEX: the Bolometer array data Analysis
software (BoA).


One reason for the need of a new software package was the new data
file format. The raw data are written by the APEX Control System
(APECS\cite{muders2006}) in MB-FITS (Multi-Beam FITS) format.
This format has been designed and constantly extended during the
first few years of APEX operations.
In addition, one of the main drivers for the development of BoA
has been to provide high modularity, and an environment where
the user can run existing scripts, but also develop his/her own
scripts to extend the capabilities of the software. As such, BoA
has been written in a scripting language, with the aim that each
processing function can easily be modified for specific needs.

BoA is an open source software. It is distributed under the terms of
the GNU General Public License. The source code and the external
libraries required to run BoA can be downloaded from a dedicated
page at the MPIfR
website\footnote{\url{http://www.mpifr-bonn.mpg.de/div/submmtech/software/boa/boa_main.html}}.
The software comes with a user and reference manual, as well as
a set of typical reduction scripts for LABOCA and SABOCA.
In the present paper, I will describe the general structure of
BoA, and highlight its main functionalities. A typical pipeline
for reducing LABOCA data is discussed in Section~\ref{sec-offline}.

\section{Structure of the software}

\subsection{Programming language}

The BoA software is written in object oriented languages: most of the
code is in Python, including many external packages, while the
most computing demanding tasks are written in FORTRAN95.

The BoA source code is made of several modules, that define
functionalities for: raw data reading, analysis tools,
plotting procedures, and data processing. The latter
applies to a so-called {\em data} object (see below),
which is created when BoA is started. More than one {\em data}
object can reside at a given time in BoA (with different variable
names). They can be dumped to binary files, in a format specific
to python, and restored later, e.g.~during the next BoA session.

\subsection{The main {\it data} object} \label{sec-data-object}

The main {\em data} object in BoA contains everything that may
be required to process one or more data file(s). The raw data
are written by the APECS\cite{muders2006}
in MB-FITS (Multi-Beam FITS) format. Once such a data-file is
read in BoA, the {\em data} object is filled up with all relevant
information, which can be divided in three groups: the telescope
(coordinates) data, the description of the bolometer array, and
the data values themselves.

\subsubsection{Telescope data}

All the information related with coordinates and time is stored into
the {\em ScanParam} attribute of the {\em data} object. This includes:
coordinates where the telescope has been pointing, in horizontal,
equatorial and galactic systems; time stamps in MJD and LST; target
name and reference coordinates; sub-scans related information: sub-scans
numbers, and integrations numbers they correspond to; when relevant,
wobbler phases. Note that the telescope positions expressed as offsets
(with respect to the reference position) in horizontal system, as they
are stored in the raw data file, include the wobbler displacement,
when relevant.

In addition, an array of flags is attached to the {\em ScanParam}
attribute, in order to flag the data recorded by all bolometers at
given times during the scan. Functions exist to flag the data in
a given interval of time, or according to the position of the
telescope (e.g. when the offset in Azimuth exceeds a given value).
It is also possible to temporarily flag (or mask) some part of the
data, and unflag it later.

\subsubsection{Bolometer array}

The most important information about the instrument in use is
stored in the MB-FITS files: telescope name and diameter; effective
observing frequency; global gain settings or attenuation factors;
for an array of detectors, relative positions of the pixels, and
their relative gains; reference channel number.
These data are
read in by BoA and stored in the {\em BolometerArray} attribute.
At the time of reading, BoA also derives the nominal instrumental
beam size from the telescope diameter and effective frequency.

The relative offsets between pixels and their relative gains can
latter be updated in BoA, by reading a so-called RCP (Receiver
Channel Parameter) file (see also Sect.~\ref{sec-rcp}).
The values written in the MB-FITS file are the ones known to APECS
at the time of observing. These may be not up-to-date, but beam maps
are observed and processed regularly, in order to update the
pixels positions and gains {\it a posteriori}.

Here also, an array of flags is associated with the {\em BolometerArray}
attribute, to allow the user to flag all data recorded by a given
pixel (or by a list of pixels).

\subsubsection{Data values and weights} \label{sec-data-data}

Finally, the data values are stored in a 2-dimensional array: the
{\em Data} attribute, with size equal to number of pixels $\times$ number
of integrations. At the time of reading, the values present in the
file are corrected for global gain/attenuation factors. They can
later be converted from instrumental units (counts) to physical
units (Jy/beam), using the conversion factor that has been measured.
They can also be corrected for the atmospheric attenuation, computed
as exp($\tau_z$ / sin(El)), where $\tau_z$ is the zenith opacity
and El is the median of the elevation during the observation.

During the reduction of an observation, the content of the {\em Data}
array is permanently updated by any processing task. For example,
applying a flat field divides all the data values for a given pixel
by the associated relative gain. A description of the main processing
tasks is given in Sect.~\ref{sec-offline}.

A second array, with the same size as the {\em Data} array, contains
the weights associated to the data. The user needs to explicitly
tell BoA to compute these weights before building a map
(Sect.~\ref{sec-map-making}). Two methods exist to
compute the weights: either using natural weighting (where each pixel
has a constant weight along the time, equal to 1/r.m.s.$^2$), or using
sliding weights, also equal to 1/r.m.s.$^2$, but this time computing the
r.m.s. within a sliding window containing a given number of integrations.
In the latter case, if all data points are flagged within a window
(which can happen e.g.~when masking a source), the corresponding
weights are computed by interpolating between available values
before and after that window.

Another 2-dimensional array contains flags, associated to individual
data values. Again, it is possible to temporarily mask part of the
data and unflag them later.

Finally, one attribute is attached to the main {\em data} object in order
to compute and store a map (Sect.~\ref{sec-map-making}).
This contains the map itself (a 2-dimensional array of pixel values),
as well as a description of the map coordinates:  projection system,
pixel scale and limits of the map.



\subsection{Associated methods}

Functions have been defined to perform all kind of data processing;
examples are described with more details in Sect.~\ref{sec-offline}.
Shortcuts have been defined, which apply to the default
{\em data} object, to ease the use of these functions from the
interactive prompt in python.
The functionalities present in BoA include: signal plotting
(Sect.~\ref{sec-visual}); flat fielding and calibration
(Sect.~\ref{sec-calib}); data flagging and unflagging
(Sect.~\ref{sec-flag}); baseline and correlated noise subtraction
(Sect.~\ref{sec-skynoise}); filtering in the Fourier domain;
computing, displaying and exporting a map (Sect.~\ref{sec-map-making}).





\section{Data visualization} \label{sec-visual}

\begin{figure}
\centering
\rotatebox{270}{\includegraphics[width=12cm]{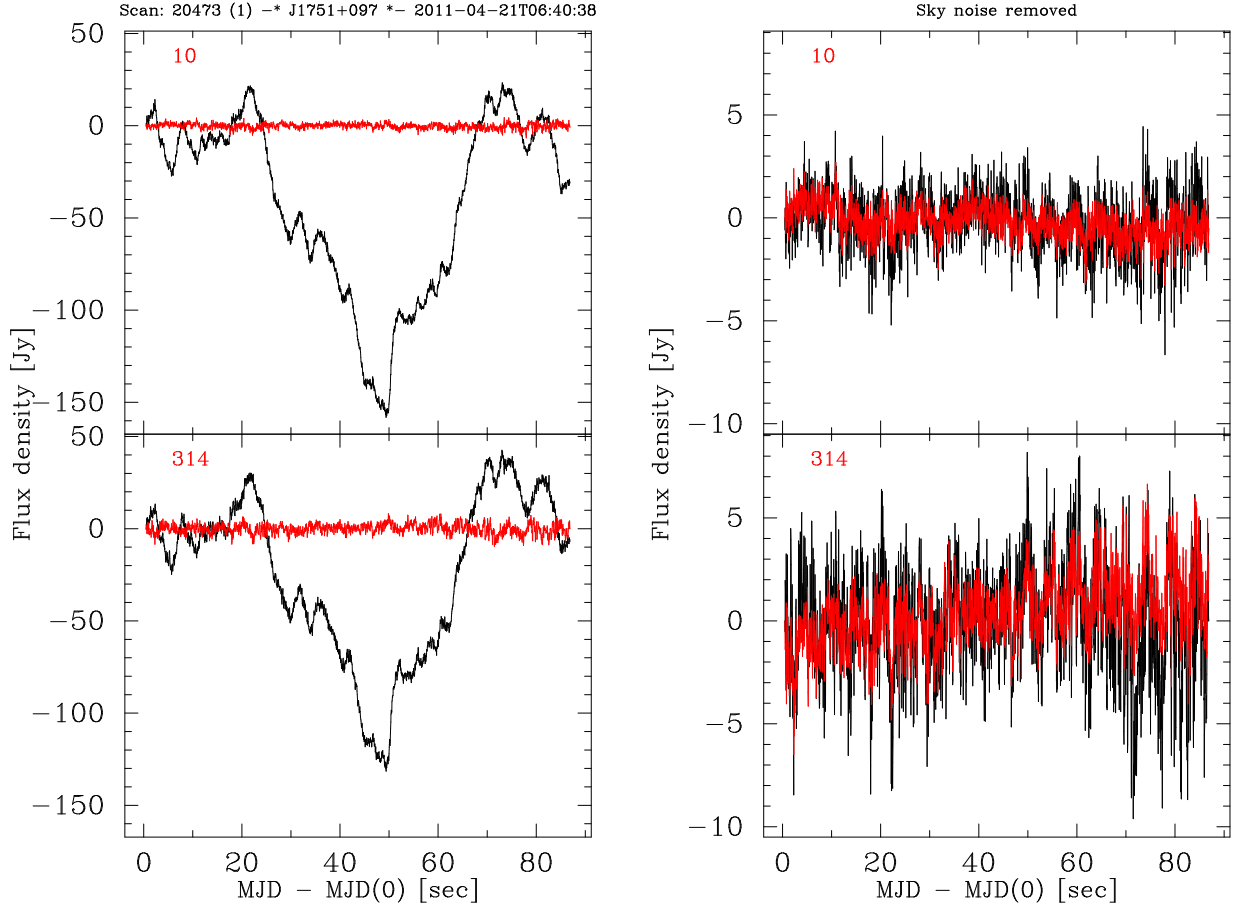}}
\vspace{1mm}
\caption{Example plots produced with the {\em signal} function, for
two bolometers. {\it Left}: raw data (black) and data after median
noise subtraction on the full array (red). {\it Right}: data after
median noise subtraction on the full array (black) and then after
median noise subtraction on amplifier boxes (red) - see also
Fig.~\ref{fig-sequence}.}
\label{fig-signal}
\end{figure}

One of the strengths of BoA is to allow the user to visualize the data
in many different ways This includes: plotting the bolometer signal vs. time
(see an example in Fig.~\ref{fig-signal}), displaying correlation plots
(signal in each bolometer vs. signal in an arbitrary reference bolometer),
plotting the mean or r.m.s. values per bolometer, computing and displaying
Fourier transform of the signals. Each of these functions can display the
data of all bolometers, or restricted to any number of bolometers. 
The user can also overlay several plots on each other, as illustrated
in Fig.~\ref{fig-signal}.

Once a map has been computed, the user can display the map itself, but
also the coverage plane or the weight plane associated to it. BoA also
provides a function to zoom into a sub-region of the map. Finally,
another function allows to look at the values of individual pixels,
or average values over a small region.
However, BoA is not aimed at providing extensive analysis tools. Once
a map has been computed (eventually co-adding several scans into one map),
the user should export the result is standard FITS format, in order to use
any other software to perform further analysis.

\section{Online data processing}

The BoA software is interfaced with the APEX Online Calibrator, which is
part of APECS. The purpose of this calibrator is to automatically perform
a quick reduction of almost all observations, and to provide feedback
to the observer. This is of particular importance for the reduction of
pointing and focus scans, since any deviation from the optimal pointing
and focus settings that can be inferred from these scans should be
corrected for before starting the next science observation.

\subsection{Pointing scans} \label{sec-pointing}

\begin{figure}
\centering
\rotatebox{270}{\includegraphics[width=11cm]{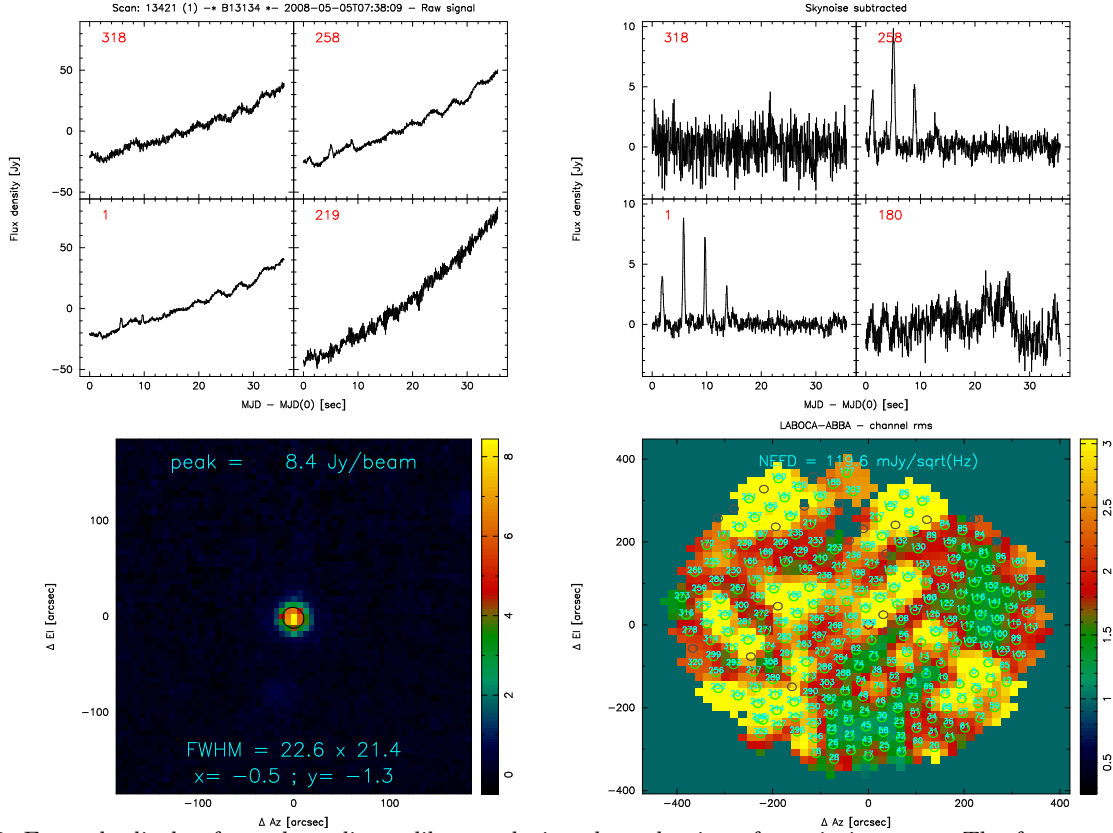}}
\caption{Example display from the online calibrator during the reduction
of a pointing scan. The four panels show: signals in the reference
pixel and three comparison pixels before any processing (top left).
Same after correlated noise has been removed (top right). Map in
horizontal coordinates built by combining all pixels (bottom left);
the results of a 2D Gaussian fit are also shown: peak flux, position
and size of the Gaussian profile. Finally, a map of the r.m.s. per pixel
is shown (bottom right), where all used pixels are plotted at their
relative positions. A quick estimate of the median value of the noise
equivalent flux densities in all channels is computed and shown on top of
this map.}
\label{fig-ex-point}
\end{figure}

LABOCA is operated in total-power mode, which means that, usually, no 
chopping secondary (wobbler) is used when observing with LABOCA.
Thanks to its high sensitivity, pointing scans on bright sources
($>$10~Jy peak flux) can be performed with a single $\sim$30~s long
spiral pattern.
Such scans are automatically recognized as pointing scans by the
online calibrator. They are processed using a quick reduction
procedure: a zero-order (median) baseline is subtracted in each
pixel; then pixels with excessively low r.m.s. ~(dead pixels, or
not seeing the sky) and the ones with very high r.m.s. ~are
flagged; then, the correlated noise is computed and subtracted
using median estimates (see Sect.~\ref{sec-skynoise} and
Fig.~\ref{fig-sequence} for details);
then, a map is computed, in horizontal coordinates (offsets
in azimuth and elevation with respect to the nominal target
position), using natural weighting. Finally, a 2-dimensional
Gaussian profile is fitted on the map, in order to derive the
observed position of the source.

The pointing errors ($\delta$Az, $\delta$El) derived from this fit are
sent by BoA to APECS. The observer can then apply these corrections.
During the reduction, the signals in the reference pixel and in three
comparison pixels are plotted, as well as the map built from the
processed data, together with the results of the Gaussian fitting
(Fig.~\ref{fig-ex-point}). Thus, the
observer can verify that the pointing offsets computed by BoA make
sense, given the quality of the map.

Correcting for the telescope pointing errors is best done as close as
possible to the science target. But the sky does not contain {\em bright}
pointing sources in all directions. Therefore, it is usual to
do such a pointing scan on a relatively faint ($\sim$1~Jy) compact
source. Then, the observer can use an existing script optimized
to process observations of such moderately faint sources.
The main difference
compared to the online reduction of pointing scans is that more
iterations of correlated noise removal are computed, in order
to decrease the noise level in the final map. Then a 2-dimensional
Gaussian profile is fitted on the map, and, if the source is
detected with a sufficient signal-to-noise, the observer can
apply the pointing corrections computed by this BoA script.

\subsection{Focus scans}

Focus scans have to be observed from time to time, in order to check
the focus settings of the telescope, and correct for them if any deviation
from the optimal position is found. This is especially important when
the air temperature changes quickly, e.g.~after sunrise or sunset.
A focus scan consists of a number of sub-scans, where each sub-scan
is a short integration (5 to 10~s) staring at a very bright source
(typically the planets Venus, Saturn or Mars, the fainter planets
Uranus and Neptune can also be used when the atmosphere is stable),
at a given focus setting. One of the focus axis (X, Y or Z) is changed
by a small amount between each sub-scan.

The online calibrator also recognizes such scans, and they are
automatically processed using a quick procedure: first, a constant
is subtracted from the signals in all pixels, so that they all
start at zero; then, a first attempt to determine the optimal
focus setting is done, by fitting a parabola to the flux observed
by the reference pixel (the one that stares on the source) as
a function of the focus position; then, the correlated noise
is estimated and subtracted from all pixels; finally, a second
parabola fit is performed to derive the optimal focus position.


Here also, the result is sent to APECS, and the observer can
apply the correction computed by BoA. The signals seen by the
reference pixel, plus three comparison pixels, are shown before
and after sky-noise removal, as well as the results of the parabola
fits. 
Thus, the observer can appreciate the
quality of the fit, and of the sky-noise subtraction.



\subsection{Science data}

Most science observations conducted with LABOCA consist in maps,
acquired with fast scanning patterns (either linear on-the-fly
maps, or one or several spirals optimized to cover a given field;
see Sect.~8 in Siringo et al.\cite{siringo2009}).
Such observations are also quickly reduced by the online calibrator,
except for the very long integrations (more than $\sim$25~min.), which
result in very big data files that would take too long to process,
and this would slow down the full system.

For data files smaller than 100~MB (which roughly corresponds to
25~min. integration with a sampling rate of 25~Hz), the following
quick reduction is performed:
pixels with very low r.m.s. (dead or not seeing the sky) or with
excessive noise are detected and flagged; correlated noise is
computed and subtracted, with three iterations on the full array;
a first-order polynomial baseline is subtracted in each pixel;
spikes are detected (data-points at more than 10-$\sigma$, where
$\sigma$ is the r.m.s. in a given pixel; note that very bright
sources are also considered as spikes in this case), and flagged;
finally, a map is computed in equatorial coordinates, using natural
weighting of the individual pixels. This map is displayed, together
with the r.m.s. in the map, computed in the part of the map with
enough coverage to be significant (i.e.~excluding the edges of the
map, where the noise level quickly increases because of the
sparse coverage).

\section{Offline data reduction} \label{sec-offline}

\begin{figure}
\centering
\includegraphics[width=5.7cm]{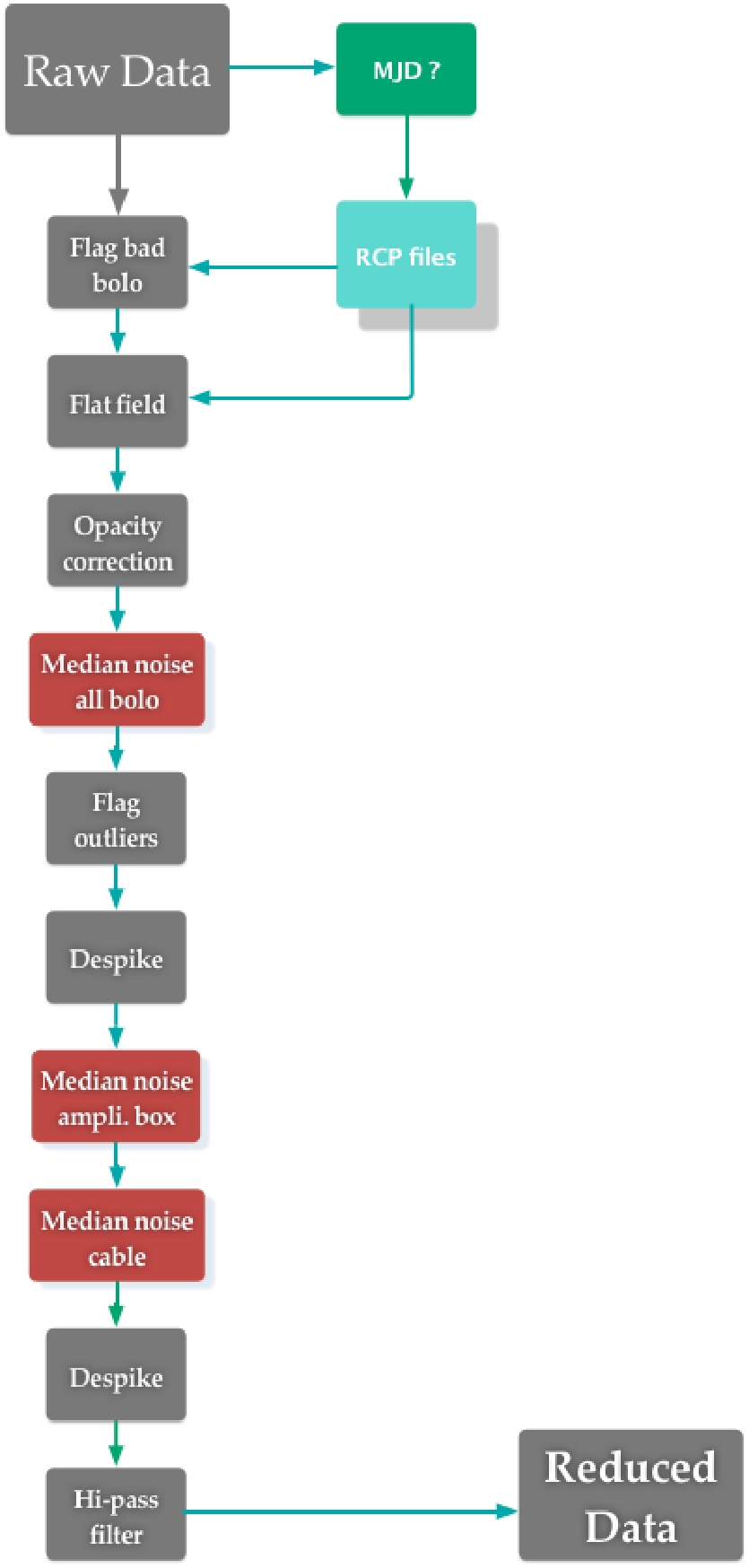}
\hspace{2.0cm}
\includegraphics[width=6.5cm]{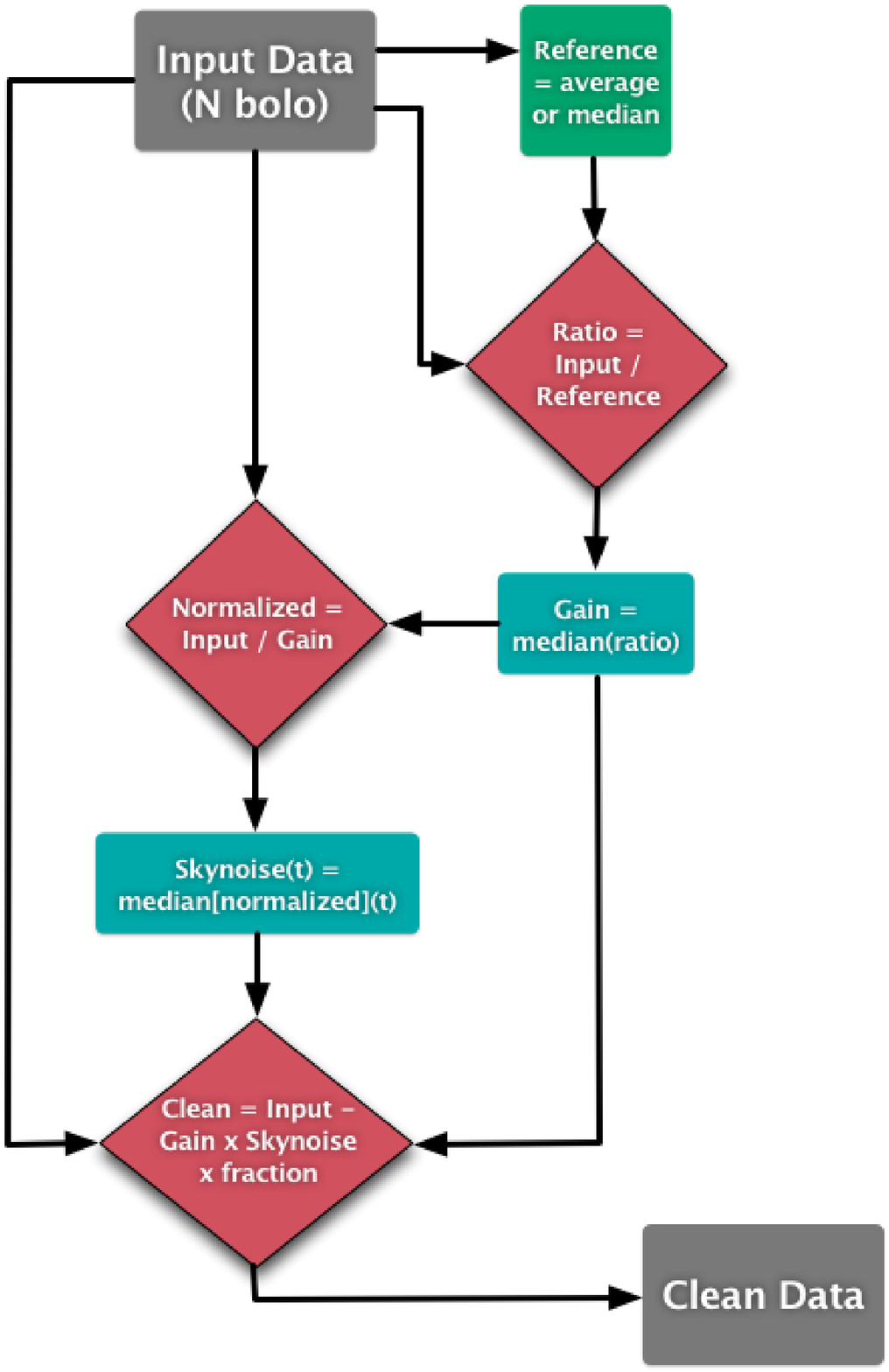}
\caption{{\it Left}: A complete reduction sequence is shown for LABOCA data,
going from the raw data to the reduced data, ready to be projected onto
a map. {\it Right}: Detailed description of the median noise subtraction
method. Not only the correlated noise, but also the relative gains are
computed using median values.}
\label{fig-sequence}
\end{figure}

In this section, a typical processing sequence data is presented.
The structure of this pipeline is illustrated in Fig.~\ref{fig-sequence}.
This applies to LABOCA as well as SABOCA data; example scripts are
provided for both instruments with the software. They cover a range
of practical cases, from faint to bright sources, and from point-like
to extended; but all scripts follow the same structure, with only
subtle differences in a few parameters, such as the number of
iterations for sky-noise subtraction, or the masking of the source.

\subsection{\label{sec-calib}Calibration}

\subsubsection{\label{sec-rcp}Updating receiver channel parameters}

Default values for the bolometer relative gains and relative positions
are contained in the raw data-file. These can be updated with more
accurate values, e.g. ~as derived from beam maps observed close in
time to the science data. BoA provides functions to select the
appropriate RCP file from a list as a function of observing time.
Then, the {\em flatfield} function should be called to divide all
bolometer signals by their relative gains.

\subsubsection{Photometric calibration}

One of the major difficulties in calibrating sub-mm data is a
proper estimate of the sky opacity. This is based on, both, regular
observations of skydips (where the bolometers are measuring the sky
emission while the telescope is tipping from high to low elevation),
and observations of primary or secondary calibrators (sources for
which the flux at the observing frequency is known). Any deviation
between the expected flux and the observed flux for these calibrators
can thus be measured, and this calibration correction factor can
be applied to the science data.

Example scripts are provided to process lists of skydips and of
calibration scans, and to store the results in text files. Then, BoA
provides functions to compute the sky opacity and the calibration
correction factor at any given time by means of linear interpolation
between the values stored in these files. This calibration should
then be applied to the science data.

Sky opacities and calibration correction factors are regularly
computed by the APEX staff, using default reduction scripts.
These values are available
online\footnote{\url{http://www.apex-telescope.org/bolometer/laboca/calibration/}}.
However, users are encouraged to process the calibration scans available with
their data themselves, in order to decide on their own the best
calibration values. Typically, a photometric uncertainty of $\sim$10\%
with LABOCA and $\sim$20\% with SABOCA can be achieved.

\subsection{\label{sec-flag}Data flagging}

Several functions have been defined to flag (or to temporarily mask)
part of the data, either within the 2D array of data values, or for
all bolometers simultaneously in given ranges of time, or to flag some
bolometers completely. For example, a function {\em despike} looks
for data values that deviate by more than a number of sigma from
the mean value in each bolometer time stream. Other functions provide
ways to mask part of the data that correspond to a certain
region of the sky, defined with a circle, or a polygon.

In addition, a function allows the user to use a map object (e.g.
~the result of a previous reduction) in order to mask in the data
the values that correspond to pixels in this map above a given
threshold (Sect.~\ref{sec-iterate}). This is very useful when bright
sources are present in the data, since their fluxes may affect some
of the processing steps, such as baseline subtraction or correlated noise
removal.

\subsection{\label{sec-skynoise}Correlated noise removal}

Because it is working in total-power mode, at a wavelength where
the emission from the atmosphere is much brighter than that from
any astronomical object (except possibly the Sun), the main source
of noise in LABOCA data comes from the slow variations of the
atmospheric emission, also called sky-noise. Fortunately, this
emission is highly correlated on the spatial scale of the detector
array. BoA provides two methods to compute (and subtract)
the correlated emission.

\subsubsection{Median noise}

The first method uses medians to compute the fraction of the signal that is
correlated between pixels. Median values are a better estimator than mean
values, because they are not affected by strong (compact) sources.
However, extended sources, with uniform emission on spatial scales
large enough so that half the pixels considered for computing the correlated
``noise'' see these sources, are considered as sky-noise and their flux
is also subtracted from the data.

By default, this procedure proceeds in two steps (see right panel
in Fig.~\ref{fig-sequence}: first, the relative gains
between pixels are estimated, using median values of the ratios between
one pixel signal and a reference signal. The reference signal can be
either the signal in one particular pixel, or the average of all signals,
or the median of all signals. For each pixel, the ratio between its signal
and the reference signal is computed as a one-dimensional array (as a
function of time or integration number), and the median value of this
ratio is considered as the relative gain of this pixel, and is stored
internally (but the signal is not modified according to this gain).
Optionally, the user can skip this step and let the function use
relative gains that were previously stored in the {\em data} object
(these gains are initialized to ones by default).

As a second step, the correlated signal is estimated, also using median
values: at each integration step, the median between all normalized
(using the relative gains computed in the previous step) signals is
considered to be the correlated noise. Finally, this correlated signal
is scaled by the relative gain of each pixel, and this (or a fraction
of this) is subtracted from all pixel signals.

This method can be applied to
all bolometers in the array, or to sub-groups of bolometers. In the
case of LABOCA, this is of particular importance to significantly
reduce the noise level: the correlated noise caused by the variations
of the sky emission is mostly correlated over the full field of view,
but additional components of correlated noise are coming from some
electronic sub-systems. In particular, groups of bolometers that
are connected to a given amplifier box (corresponding to one quarter
of the field of view), and sub-groups that are connected on a given
flat cable (corresponding to a spatial scale of $\sim$5$'$$\times$2$'$,
see Siringo et al.\cite{siringo2009} for details on the electronic scheme of LABOCA)
show correlated electronic noise. However, special care should
be taken in order not to filter out real extended emission; when
only point sources are expected, iterating several times on the
correlated noise removal by groups of pixels helps in decreasing
the final noise level.

\subsubsection{Principal component analysis}

The other correlated noise removal function is based on a principal
component analysis (PCA) and filtering.
Comparisons between the results obtained when applying this
method, and with the median noise removal, showed that they give
similar results, both with LABOCA and with APEX-SZ data.
However, the effect of PCA filtering on the true astronomical signal
is difficult to predict. Therefore, most results published so far
based on BoA data reduction were obtained using the median noise removal
method. More elaborated methods, e.g. ~fitting and removing a low-order
two-dimensional polynomial function across the array at each time step,
can also easily be implemented in BoA.

\subsection{\label{sec-map-making}Map making}

At any time during the reduction, a map can be computed, using all
or part of the data currently in memory. Three coordinate systems are
implemented in BoA: Horizontal (offsets in azimuth and elevation with
respect to the reference coordinate),
Equatorial J2000 and Galactic coordinates. The map-making procedure
computes a map by dumping and co-adding the data values on a grid of
pixels, without using extra information from the redundancy within
the map. By default, the map-making procedure dumps each data value into
the nearest pixel; optionally, the value can be divided into the four
neighbour pixels, with fractions depending on the exact location of that
data point with respect to the pixel grid. The mandatory arguments
for the mapping function are the limits in X and Y (when not given, default
values are computed in order to encompass the full region covered by the data)
and the pixel scale (default value is half the instrumental beam).
The data values are co-added using a weighted average, where the weights
have to be computed prior to building the maps (Sect.~\ref{sec-data-data}).

The resulting map can be dumped into a binary file, that can be later
imported again in BoA. It can also be exported into a FITS file, with
a standard FITS header containing a description of the coordinates
following the WCS specification\cite{ref-fits-wcs}.
Currently, the only projection system implemented in BoA is GLS (Global
Sinusoid, equivalent to the Sanson-Flamsteed - SFL - projection; see
Calabretta \& Greisen\cite{ref-fits2} for details).
By default, the output FITS file
contains three planes: the flux map, the weight map and the coverage
map. The weights are computed by summing all data weights that have
contributed to one given pixel. The coverage plane is similar, but
considering all weights equal to one (in other words, the coverage
gives the number of hits in a given pixel).

When several observations cover the same region of the sky (or if
they overlap), they can be processed independently, and maps covering
exactly the same region (and with the same pixel scale) can be built from
each observation. These maps can then be co-added, using a weighted
average, into one combined map. This ``final'' map can also be exported
as a binary (python) file, or into a FITS file.

\subsection{\label{sec-iterate}Iteration with source model}

At the end of the reduction sequence, once a map has been computed
(eventually combining several scans), it is sometimes useful to
repeat the full processing in an iterative way. The result map
can be used as a model to mask in the data the measurements
that correspond to pixels in the map above a given threshold.
It is also possible to compute a signal-to-noise map and use
it as a model to mask the data points where highly significant
emission has been found. The processing steps are then computed
ignoring the masked data, but the results (for example subtracting
a baseline) are also applied to them.

Such iterative reduction scheme is of critical importance in the presence
of bright sources, for example in the Galactic plane\cite{ref-atlasgal}.
In such complex fields, masking the data that correspond to strong
emission is not the best solution, because too much data are masked
when also extended emission becomes significant. Also, a drawback
of masking part of the data is that spikes that sit on top of a bright 
source can never be identified and flagged. For these reasons, it is
better to subtract a model (i.e. ~the map resulting from the previous
iteration) from the raw calibrated data, and then repeat the full
reduction sequence.

This iterative scheme can be repeated any number of times. For example,
in the case of the ATLASGAL\cite{ref-atlasgal} data, 15 iterations have
been performed to produce the final maps. Also when no really strong source
is present in the field, this iterative processing can be useful in
order to recover as much extended emission as possible. This method
has been used by Belloche et al.\cite{ref-belloche} on LABOCA data
covering the Chamaeleon molecular cloud. They also performed extensive
simulations in order to estimate the fraction of flux lost by the
data processing as a function of source size and of number of iterations.
The results that they present in their appendix apply to any LABOCA
data processed in a similar manner, i.e. ~with the median noise
removal method and an iterative reduction scheme.



\section{Conclusion}

In the past few years, BoA has been used by an increasing number of
people to process LABOCA, SABOCA and APEX-SZ data. The results have
been published in major journals of astrophysics research.

Since BoA is an open-source software, anybody can contribute to
its further development, either by adding functions in the source
code of BoA itself, or by writing and sharing reduction scripts.
Current developments at APEX and at the MPIfR include a parallelization
of the most demanding processing steps, in order to make BoA able to
reduce data taken with the next generation of kilo-pixels cameras.

\acknowledgments     

I want to cheerfully thank all people who contributed to the development
of BoA over the years. This includes: Marcus Albrecht, Alexandre Beelen,
Frank Bertoldi, Thomas J\"urges, Attila Kov\'acs, Dirk Muders,
Reinhold Schaaf, Martin Sommer, Catherine Vlahakis and Axel Wei\ss.
I also want to thank all the past and present operators and astronomers
at APEX, who have been using BoA from the beginning. As such, they greatly
contributed to the development and bug-fixing of BoA and of the common
reduction scripts.


\bibliography{boaref}   
\bibliographystyle{spiebib}   

\end{document}